\pdfoutput=1

\documentclass[aps,superscriptaddress,twocolumn,secnumarabic,showpacs,showkeys,amsmath,amssymb,nofootinbib,tightenlines,floatfix]{revtex4}

\usepackage{graphicx}
\usepackage{dcolumn}

\newcommand{\back}{\negmedspace}

\newcommand{\Ldbrace}{\Bigl \{ \negmedspace \Bigl \{}
\newcommand{\Rdbrace}{\Bigr \} \negmedspace \Bigr \}}
\newcommand{\Ldbbrace}{\Biggl \{ \negmedspace \Biggl \{}
\newcommand{\Rdbbrace}{\Biggr \} \negmedspace \Biggr \}}

\newcommand{\Ldbrack}{\Bigl [ \negmedspace \Bigl [}
\newcommand{\Rdbrack}{\Bigr ] \negmedspace \Bigr ]}
\newcommand{\Ldbbrack}{\Biggl [ \negmedspace \Biggl [}
\newcommand{\Rdbbrack}{\Biggr ] \negmedspace \Biggr ]}

\newcommand{\al}{\boldsymbol {\displaystyle \aleph}}
\newcommand{\bi}{\boldsymbol {\imath}}

\allowdisplaybreaks[4]
\begin{document}

\title{Combinatorial Entropies and Statistics}


\author{Robert K. Niven}
\email{r.niven@adfa.edu.au}
\affiliation{School of Aerospace, Civil and Mechanical Engineering, The University of New South Wales at ADFA, Northcott Drive, Canberra, ACT, 2600, Australia.}



\begin{abstract}
We examine the {combinatorial} or {probabilistic} definition (``Boltzmann's principle'') of the entropy or cross-entropy function $H \! \propto \! \ln \mathbb{W}$ or $D \! \propto \! - \ln \mathbb{P}$, where $\mathbb{W}$ is the statistical weight and $\mathbb{P}$ the probability of a given realization of a system. Extremisation of $H$ or $D$, subject to any constraints, thus selects the ``most probable'' (MaxProb) realization. If the system is multinomial, $D$ converges asymptotically (for number of entities $N \back \to \back \infty$) to the Kullback-Leibler cross-entropy $D_{KL}$; for equiprobable categories in a system, $H$ converges to the Shannon entropy $H_{Sh}$.  However, in many cases $\mathbb{W}$ or $\mathbb{P}$ is not multinomial and/or does not satisfy an asymptotic limit. Such systems cannot meaningfully be analysed with $D_{KL}$ or $H_{Sh}$, but can be analysed directly by MaxProb. This study reviews several examples, including (a) non-asymptotic systems; (b) systems with indistinguishable entities (quantum statistics); (c) systems with indistinguishable categories; (d) systems represented by urn models, such as ``neither independent nor identically distributed'' (ninid) sampling; and (e) systems representable in graphical form, such as decision trees and networks. Boltzmann's combinatorial definition of entropy is shown to be of greater importance for {``probabilistic inference''} than the axiomatic definition used in information theory.
\end{abstract}

\maketitle


\section{Introduction}

The {\it combinatorial} or {\it probabilistic} definition of entropy, given by Boltzmann, is usually written as \cite{Boltzmann_1877, Planck_1901}: 
\begin{equation}
S_N  = NS = k \ln \mathbb{W}  
\label{eq:Boltz1}
\end{equation}
where $S_N$ is the total thermodynamic entropy of a system, $S$ is the entropy per unit entity, $N$ is the number of entities, $\mathbb{W}$ is number of occurrences of a specified realization of the system (its statistical weight) and $k$ is the Boltzmann constant. This can be rewritten to give dimensionless forms of the entropy and cross-entropy (directed divergence or negative relative entropy) functions, respectively \cite{Vincze_1972, Grendar_G_2001, Niven_CIT, Niven_2005, Niven_2006, Niven_MaxEnt07, Niven_CTNext07, Niven_non-asymp08, Niven_G_MBBEFD_08}:
\begin{gather}
H = K \ln \mathbb{W}
\label{eq:Boltz2}
\\
D = - K \ln \mathbb{P}
\label{eq:Boltz3}
\end{gather}
where $\mathbb{P}$ is the probability of a given realization and $K$ is a dimensionless constant. Since $\ln x$ is monotonic with $x$, maximisation of $H$ or minimisation of $D$, subject to the constraints on a system, {\it always} yields its ``most probable'' (MaxProb) realization(s), and so can be used to infer the properties of the system.  If a system is governed by the multinomial weight or distribution, respectively:
\begin{gather}
\mathbb{W}_{mult} = \frac{N!}{ \prod\nolimits_{i=1}^s n_i!}
\label{eq:multW}
\\
\mathbb{P}_{mult} = N!  \prod\nolimits_{i=1}^s \frac{q_i^{n_i}}{n_i!}
\label{eq:multP}
\end{gather}
where $n_i \in \{\mathbb{N} \cup 0\}$ is the occupancy of each category $i=1,...,s$ and $q_i$ is its source (``prior'') probability, then \eqref{eq:Boltz2}-\eqref{eq:Boltz3} with $K \back = \back N^{-1}$ converge asymptotically ($N \back \to \back \infty$) \cite{Sanov_1957} to the Shannon entropy \cite{Shannon_1948} or Kullback-Leibler cross-entropy functions \cite{Kullback_L_1951, Kullback_1959}: 
\begin{gather}
H_{Sh}  = \lim\limits_{N \to \infty} \frac{1}{N} \ln \mathbb{W}_{mult} =  - \sum\limits_{i = 1}^s {p_i \ln p_i } 
\label{eq:Shannon}
\\
D_{KL} = - \lim\limits_{N \to \infty} \frac{1}{N} \ln \mathbb{P}_{mult} = \sum\limits_{i = 1}^s {p_i \ln \frac{{p_i }}{{q_i }}} 
\label{eq:KL}
\end{gather}
where $p_i=n_i/N$ is the frequency or probability of occupancy of the $i$th category. Eqs.\ \eqref{eq:Shannon}-\eqref{eq:KL} are commonly used in the maximum entropy (MaxEnt) or minimum cross-entropy (MinXEnt) extremisation methods to infer the ``least informative'' or ``most uncertain'' distribution $p_i^*$ of the system \cite{Jaynes_1957, Jaynes_1963, Kapur_K_1992, Jaynes_2003}, based on axiomatic justifications developed in information theory \cite{Shannon_1948, Shore_J_1980}. 

It is important to recognise, however, that $\mathbb{W}$ or $\mathbb{P}$ may not be multinomial and/or may not satisfy an asymptotic limit. Extremisation methods based on \eqref{eq:Shannon} or \eqref{eq:KL} will then give a distribution which is unrepresentative of the system, except in special instances.  In such cases, it is preferable to apply the MaxProb principle \eqref{eq:Boltz2}-\eqref{eq:Boltz3} directly, to obtain the most probable distribution of the system. Of course, it is recognised that in non-asymptotic systems ($N \back \ll \back \infty$), the most probable distribution may not be the only observable distribution; in other words, there may be a significant spread around the inferred distribution \cite{Niven_non-asymp08}.  Furthermore, due to quantisation effects, the actual realizable MaxProb distribution(s) may be sub-optimal \cite{Niven_non-asymp08}. Despite these effects, the MaxProb principle provides a powerful tool for ``probabilistic inference'' of the properties of a probabilistic system, irrespective of its form.

The aim of this work is to demonstrate the utility of the MaxProb principle \eqref{eq:Boltz2}-\eqref{eq:Boltz3} in a number of systems of physical interest:\ (a) non-asymptotic systems; (b) systems with indistinguishable entities (quantum statistics); (c) systems with indistinguishable categories; (d) systems represented by urn models, e.g.\ ``neither independent nor identically distributed'' (ninid) sampling; and (e) systems representable in graphical form, such as decision trees and networks. Definitions of terms are provided in \S\ref{Defs}, following which the above systems are examined in \S\ref{Mult}-\ref{Graphs}.  Particular attention is paid to (c), to explore the peculiar properties of systems with indistinguishable categories.  The case studies serve as evidence that Boltzmann's definition \eqref{eq:Boltz2}-\eqref{eq:Boltz3} is of much greater utility for {probabilistic inference} than the Shannon or Kullback-Leibler functions \eqref{eq:Shannon}-\eqref{eq:KL} of information theory.

\section{\label{Defs}Definitions}
\begin{figure}[h]
\setlength{\unitlength}{0.6pt}
  \begin{picture}(450,205)
   \put(0,0){\includegraphics[width=85mm]{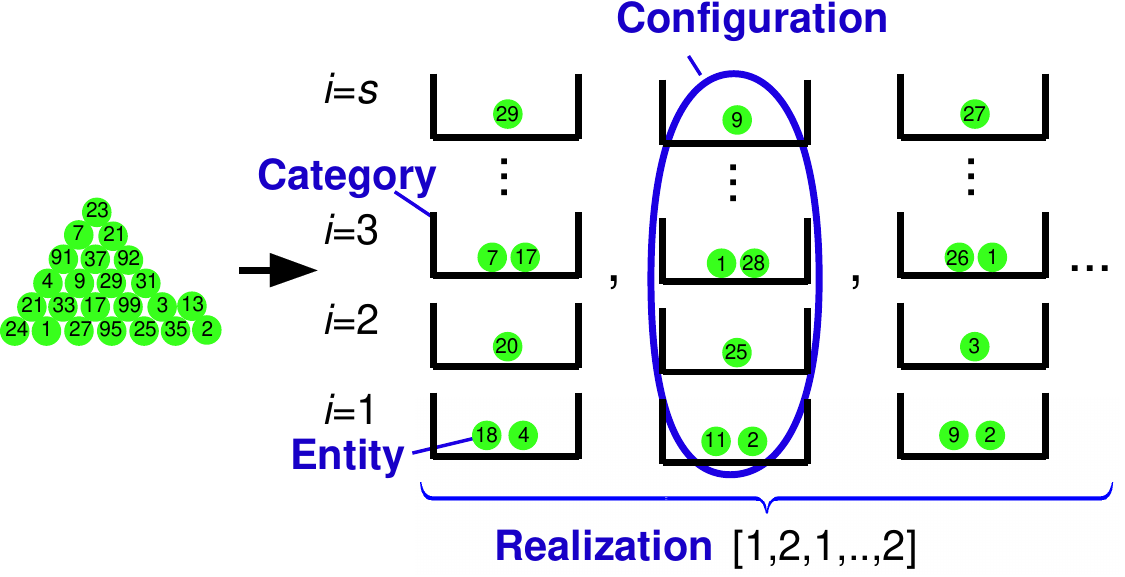} }
  \end{picture}
\caption{Definition of terms used in the combinatorial definition of entropy.}
\label{fig:defs}
\end{figure}

To avoid confusion, it is necessary to define several terms, discussed in reference to the combinatorial allocation scheme (``ball-in-box'' model) shown in Figure \ref{fig:defs} \cite{Niven_CIT, Niven_2005, Niven_2006, Niven_MaxEnt07, Niven_CTNext07, Niven_non-asymp08, Niven_G_MBBEFD_08}; this scheme encompasses both physical and mathematical (information-theoretic) interpretations.  We make the following definitions:
\begin{list}{$\bullet$}{\topsep 2pt \itemsep 2pt \parsep 0pt \leftmargin 8pt \rightmargin 0pt \listparindent 0pt
\itemindent 0pt}
\item An {\it entity} $u_m, m=1,...,N$ is a discrete particle, object or agent, or an individual selection of a discrete random variable, which acts separately but not necessarily independently of other entities.  
\item A {\it category} $c_i, i=1,...,s$ is a possible assignment of an entity (e.g.\ an energy level, side of a die or alphabetic symbol). Although not shown in Figure \ref{fig:defs}, categories can be {\it degenerate} (involving $g_i$ subcategories in each category $i$) and/or {\it multivariate} (involving a vector index $\bi$).
\item A {\it probabilistic system} is the ordered triple $\Upsilon (U,C,\Psi)$, consisting of a finite set of entities $U=\{u_m\}$; a finite set of categories $C=\{c_{\bi}\}$ (possibly a set of multivariate degenerate sets) with $C \cap U = \varnothing$; and a discrete random variable $\Psi: U \! \to \! C$. In other words, $\Psi$ is a function which assigns all entities $u_m \in U$ to selected categories $c_{\bi} \in C$ in accordance with some probabilistic rule (not all categories need be selected). This definition encompasses both physical and mathematical situations.
\item A {\it configuration} is an identifiable permutation or pattern of entities amongst the categories, i.e.\ a set of assignments $\{U \! \to \! C\}$ (in physics, a {\it complexion} or {\it microstate}; in gambling or informatics, a {\it sequence}). A configuration is thus a property of a system as a whole.
\item A {\it realization} is an aggregated arrangement of entities amongst the categories of a system, i.e.\ a set of configurations $\{\{ U \! \to \! C\}^{(1)}, \{ U \! \to \! C\}^{(2)}, ...\}$, as specified by some rule. A common rule is to take the number of entities in each category, as specified by the occupancy vector or tensor ${\bf n} = \{n_{\bi}\}$ (in physics, a {\it macrostate}; in informatics, an {\it outcome} or {\it type}). Realizations are here considered mutually exclusive (this requirement could be relaxed to give some very different types of systems).
\item The {\it statistical weight} $\mathbb{W}^{(\nu)}$ of the $\nu$th realization ${\bf n}^{(\nu)}$ is the number of ways in which it can occur, i.e.\ its number of configurations.
\item The {\it governing probability} $\mathbb{P}^{(\nu)}$ of the $\nu$th realization ${\bf n}^{(\nu)}$ is its probability of occurrence, i.e.\ the sum of probabilities of its component configurations.
\end{list}
Figure \ref{fig:defs} shows the allocation of distinguishable entities to distinguishable categories, without replacement, until all $N$ available entities are exhausted (see \S\ref{Mult}). This allocation scheme can be varied in many ways.

We therefore wish to conduct {\it probabilistic inference}, i.e.\ to infer the properties of a probabilistic system $\Upsilon (U,C,\Psi)$, using the available information about its set of realizations $\{ {\bf n}^{(\nu)}\}$ with weights $\{ \mathbb{W}^{(\nu)}\}$ or probabilities $\{ \mathbb{P}^{(\nu)}\}$.  Two ``measures of central tendency'' are evident:
\begin{list}{$\bullet$}{\topsep 3pt \itemsep 3pt \parsep 0pt \leftmargin 8pt \rightmargin 0pt \listparindent 0pt
\itemindent 0pt}
\item One measure - arguably the most important for inferring the ``typical'' behaviour of a system - is the {\it most probable} (MaxProb) or {\it modal} realization \cite{Boltzmann_1877, Planck_1901, Vincze_1972, Grendar_G_2001, Niven_CIT, Niven_2005, Niven_2006, Niven_MaxEnt07, Niven_CTNext07, Niven_non-asymp08, Niven_G_MBBEFD_08}:
\begin{equation}
{\bf n}^\# = \arg \sup\limits_{\nu} \mathbb{W}^{(\nu)} = \arg \sup\limits_{\nu} \mathbb{P}^{(\nu)}
\label{eq:MaxProb}
\end{equation}
Its use depends on the principle that {\it ``A system can be represented by its realization of highest probability''}. A significant advantage of MaxProb is that it can often be found by extremisation or optimisation methods. Of course, multimodal distributions will have multiple maxima, an inherent aspect of this method \cite{Grendar_G_2001, Niven_CIT, Niven_MaxEnt07}.
\item Another measure is the mean-weighted, superpositional or expected occurrence realization (MeanProb), in which each realization is weighted by its weight or probability \cite{Grendar_G_2001}:
\begin{gather}
\overline{\bf n} = \frac{ \sum\limits_{\nu} {\bf n}^{(\nu)} \mathbb{W}^{(\nu)} } { \sum\limits_{\nu} \mathbb{W}^{(\nu)} } 
= { \sum\limits_{\nu} {\bf n}^{(\nu)} \mathbb{P}^{(\nu)} }  
\label{eq:MeanProb}
\end{gather}
This measure is important for non-asymptotic systems and those with skewed distributions, but its calculation can become formidable as the number of realizations increases (often, an exponential function of $N$).  
\end{list}
Both MaxProb and MeanProb are independent of any information-theoretic or axiomatic considerations, other than those of probability theory itself. This is absolutely essential, since in any contradiction between information theory and probability theory, the latter - being more fundamental - must triumph \cite{Niven_MaxEnt07}. The two measures also do not require asymptotic behaviour, and so can be applied to systems with finite numbers of entities \cite{Niven_2005, Niven_2006, Niven_MaxEnt07, Niven_CTNext07, Niven_non-asymp08, Niven_G_MBBEFD_08}.

Whilst this study contains distinct philosophical differences with Jaynes \cite{Jaynes_1957, Jaynes_1963, Jaynes_2003} over the philosophical meaning of the entropy concept, the ``subjective Bayesian'' definition of probabilities - as assignments based on what we know - is adopted here. It is also recognised that there are many different ways to assign entities and categories within a system, and many ways to group configurations into realizations, with any particular choice being dependent on the observer's {\it purpose}.  This leads to the ``subjective'' (or ``observer-dependent'') interpretation of the entropy concept, a viewpoint staunchly defended by Jaynes \cite{Jaynes_1957}. This was aptly expressed by Tseng and Caticha \cite{Tseng_C_2002}:
\begin{list}{~}{\topsep 6pt \itemsep 0pt \parsep 0pt \leftmargin 8pt \rightmargin 0pt \listparindent 0pt
\itemindent 0pt}
\item {\it ``Entropy is not a property of a system $\dots$ [it] is a property of our description of a system.''}
\end{list}
Different observers (indeed, the same observer), with different available information and/or different purposes, can therefore make different probability and entropy assignments for the same system, leading to different (rational) conclusions; this is a necessary feature of probabilistic inference. The test of validity of such inference is the extent of its agreement with observations, responsibility for which again lies with the observer and his/her social cohort. Such sentiments in no way weaken the mathematical rigour of the probabilistic method, as set out in the following sections, nor the rules of probability theory upon which it is based. 


\section{\label{Mult}Non-Asymptotic Multinomial Systems}
We first examine univariate multinomial systems, the original application of Boltzmann's principle \cite{Boltzmann_1877, Planck_1901}. From a Bayesian perspective, there are many reasons why one might (rationally) select the multinomial distribution \eqref{eq:multP} to represent a system \cite{Niven_MaxEnt07}; it encompasses, but does not imply, a ``frequentist'' approach \cite{Jaynes_1957, Jaynes_1963, Jaynes_2003}. For maximum  generality, we include the source or prior distribution $q_i$; in physics, this is often interpreted as the number of distinguishable subcategories or {\it degeneracy} $g_i$ of each category $i$, normalised by the total degeneracy of the system $G=\sum\nolimits_{i=1}^s g_i$ \cite{Niven_G_MBBEFD_08}. 
For constant $N$, applying the combinatorial definition \eqref{eq:Boltz3} to the multinomial distribution \eqref{eq:multP} (taking $K=N^{-1}$) yields the non-asymptotic cross-entropy function \cite{Niven_2005, Niven_2006, Niven_MaxEnt07, Niven_non-asymp08, Niven_G_MBBEFD_08}:
\begin{equation}
-D_{mult}^{(N)}=\frac{1}{N} \ln \mathbb{P}_{mult} = \frac{1}{N} \Bigr\{ \ln N! + \sum\limits_{i = 1}^s \bigr[ n_i \ln q_i  - \ln n_i ! \bigr] \Bigr\}.
\label{eq:Dx}
\end{equation}
Either \eqref{eq:Dx}, or $\ln \mathbb{P}_{mult}$ itself, can be maximised by the Lagrangian method subject to the constraints:
\begin{gather}
\sum\limits_{i = 1}^s n_i = N,
\label{eq:C0}
\\
\sum\limits_{i = 1}^s n_i f_{ri} = F_r, \qquad r=1,...,R. 
\label{eq:C1}
\end{gather}
where $f_{ri}$ is some function of each category $i$ and $F_r$ is its total value, to infer the most probable distribution of the system \cite{Niven_2005, Niven_2006, Niven_non-asymp08, Niven_G_MBBEFD_08}:
\begin{equation}
p_i^{\#} = \frac{n_i^{\#}}{N} = \frac{1}{N} \, \Lambda^{-1} \Bigl[ \frac{1}{N} \ln N! + \ln q_i - \lambda_0 - \sum\limits_{r=1}^R \lambda_r f_{ri} \Bigr]  
\label{eq:phash}
\end{equation}
where $\lambda_r$ is the Lagrangian multiplier associated with the $r$th constraint, $\Lambda^{-1}(y)= \psi^{-1}(y-1)$ is the upper inverse of the function $\Lambda(x)=\psi(x+1)$ and $\psi(x)$ is the digamma function. The Massieu function $\lambda_0$ cannot be factored from \eqref{eq:phash}, hence the latter must be solved simultaneously with all constraints \eqref{eq:C0}-\eqref{eq:C1}.  In the asymptotic limit $N \to \infty$, the above extremisation converges to the Boltzmann distribution:
\begin{equation}
\begin{split}
p_i^{*} = \frac{n_i^{*}}{N} &= q_i \exp ( - \lambda_0' - \sum\limits_{r=1}^R\lambda_r f_{ri} ) 
\\
&= Z^{-1}{ q_i \exp (- \sum\limits_{r=1}^R \lambda_r f_{ri} ) } 
\end{split}
\label{eq:pstar}
\end{equation}
where $\lambda_0'=\lambda_0+1$ and $Z={\sum\nolimits_{i=1}^s  q_i \exp (- \sum\nolimits_{r=1}^R \lambda_r f_{ri} ) }$ is the partition function.  

The effect of $N$ on the properties of non-asymptotic multinomial systems, including (i) the discrepancy between inferred MaxProb and Kullback-Leibler MinXEnt distributions \eqref{eq:phash}-\eqref{eq:pstar}, (ii) the spread of realizations around the MaxProb distribution, and (iii) the importance of quantisation, are examined elsewhere \cite{Niven_non-asymp08}. The analyses reveal the importance of $N$ in statistical mechanics.  The information-theoretic properties of non-asymptotic multinomial systems have also been examined, in which the change in ``information'' is defined as the negative change in the non-asymptotic entropy analogue of \eqref{eq:Dx} \citep[c.f.][]{Szilard_1929, Brillouin_1950, Brillouin_1951a, Brillouin_1953, Tribus_M_1971}, i.e.:
\begin{equation}
\Delta I \text{ (bits)}= - \frac{\Delta H}{\ln 2} 
= - \frac {K \Delta \ln \mathbb{W}}{\ln 2}  
\label{eq:infm}
\end{equation}
both for binary systems ($s=2$) \cite{Niven_2005} and equiprobable systems in general \cite{Niven_2006}. The analyses show that ``information'' consists of two parts: one associated with knowledge of the realization $\{n_i\}$ and the other associated with knowledge of $N$. Such findings overturn the prevailing wisdom in communications and information theory, in which $N$ is assumed to be infinite and therefore irrelevant \cite{Shannon_1948, Shore_J_1980}. 

The MaxProb principle has also been applied to the analysis of a non-asymptotic, closed thermodynamic system of non-interacting particles (a double system-bath with heat transfer), using the multinomial distribution \cite{Niven_non-asymp08}. This shows that in such systems, thermodynamic intensive variables such as temperature are well-defined at small $N$ and do not require a ``thermodynamic limit'' \cite{Niven_non-asymp08}.  This concurs with similar findings by other workers, from different perspectives \cite{Gross_2001}, as well as with common practice in engineering analyses of heat transfer \cite{Bird_etal_2006}. 

\section{\label{Indisting_Ent}Distinguishability of Entities}

Consideration of the effect of indistinguishable entities in the 1920s is perhaps the most famous application of MaxProb \cite{Bose_1924, Einstein_1924, Einstein_1925, Fermi_1926, Dirac_1926}, providing the groundwork for the development of quantum theory.  This has now led to the following four allocation schemes (in physics, referred to as ``statistics'') \cite{Bose_1924, Einstein_1924, Einstein_1925, Fermi_1926, Dirac_1926, Brillouin_1927, Brillouin_1930, Tolman_1938, Davidson_1962, Lynden-Bell_1967, Arad_LB_2005, Bindoni_S_2008}:
\begin{list}{$\bullet$}{\topsep 2pt \itemsep 2pt \parsep 0pt \leftmargin 8pt \rightmargin 0pt \listparindent 0pt \itemindent 0pt}
\item {\it Maxwell-Boltzmann} (MB) statistics, in which distinguishable entities are allocated to distinguishable degenerate categories, with no restrictions on the occupancies;
\item {\it Lynden-Bell} (LB) statistics, as for Maxwell-Boltzmann statistics but with a maximum of one entity per subcategory \cite{Lynden-Bell_1967, Arad_LB_2005, Bindoni_S_2008}. 
\item {\it Bose-Einstein} (BE) statistics, in which indistinguishable entities ({\it bosons}) are allocated to distinguishable degenerate categories, with no restrictions on the occupancies; and
\item {\it Fermi-Dirac} (FD) statistics, as for Bose-Einstein statistics (involving {\it fermions}) but with a maximum of one entity per subcategory.
\end{list}
BE and FD statistics were developed for quantum systems, but have found many other applications, e.g. the application of FD statistics to the packing of granular materials \cite{Ghaderi_2006}.  LB statistics were developed for collisionless particle systems, such as gravitational stellar dynamics \cite{Lynden-Bell_1967, Arad_LB_2005, Bindoni_S_2008}.  The commonly adopted statistical weights of these statistics are given in Table \ref{tab:disting} \citep[e.g.][]{Bose_1924, Einstein_1924, Einstein_1925, Fermi_1926, Dirac_1926, Brillouin_1927, Brillouin_1930, Tolman_1938, Davidson_1962, Lynden-Bell_1967, Arad_LB_2005, Bindoni_S_2008}. Note that the MB statistic is multinomial \eqref{eq:multP}.  
Only the simplest, univariate version of each statistic is given here; their formulation is scrutinised more closely in \cite{Niven_G_MBBEFD_08}.  

From the combinatorial definition of entropy \eqref{eq:Boltz2} and MaxProb principle \eqref{eq:MaxProb}, the non-asymptotic and asymptotic entropy functions and most probable distributions - calculated subject to the constraints \eqref{eq:C0}-\eqref{eq:C1} - are listed in Table \ref{tab:disting}.  As with multinomial systems (\S\ref{Mult}), the inferred non-asymptotic most probable distribution obtained by extremisation may differ from the actual (realizable) distribution(s), due to quantisation effects \cite{Niven_non-asymp08}. Note that the asymptotic LB and FD distributions are identical up to normalisation, although their meaning is different \cite{Lynden-Bell_1967, Arad_LB_2005, Bindoni_S_2008}. The BE and FD weights converge to $\mathbb{W}_{MB}/N!$ in highly degenerate systems $g_i \back \gg \back n_i$, whilst the LB weight converges directly to $\mathbb{W}_{MB}$; in the same limit, the LB, BE and FD entropies and most probable distributions all converge (up to a constant) to those of the MB distribution. From the pattern of the weights (Table \ref{tab:disting}), we can also define a distinguishable-entity equivalent of BE statistics with weight $\mathbb{W}_{D:D}  = N! \prod\nolimits_{i = 1}^s {{{(g_i  + n_i  - 1)!}}/{{(g_i  - 1)! \, n_i !}}}$, which for $g_i \back \gg \back n_i$ also converges to $\mathbb{W}_{MB}$; this does not appear to have been examined previously.

The non-asymptotic BE and FD statistics have important information-theoretic implications \cite{Niven_2005, Niven_2006}.  Using the combinatorial definition of information \eqref{eq:infm}, it is shown that the observation of a finite number of bosons or fermions requires the input of energy or information; from the second law of thermodynamics, this is thermodynamically irreversible. A single boson or fermion must therefore appear to behave as if it were an infinite number of entities until its moment of observation. This ``information relativity'' perspective provides a rational explanation for the ``collapse of the wavefunction'' in quantum systems, which is not explained by present-day quantum theory, and for which many metaphysical justifications have been proposed \cite{Feynman_1963}. 

It is also possible to derive {\it intermediate} statistics which interpolate between BE and FD statistics. Several alternatives are available:
\begin{list}{$\bullet$}{\topsep 2pt \itemsep 2pt \parsep 0pt \leftmargin 8pt \rightmargin 0pt \listparindent 0pt \itemindent 0pt}
\item {\it Gentile statistics}, which indistinguishable entities are allocated to distinguishable categories with restriction $n_i \in \{0,1,...,m\}$ entities per subcategory \cite{Gentile_1940, Guenault_M_1962, Kapur_K_1992, Hernandez-Perez_T_2007}.
\item {\it Haldane-Wu statistics}, in which entities are allocated to categories using a generalised Pauli exclusion principle \cite{Haldane_1991, Wu_1994}.
\item {\it Acharya-Swamy statistics}, proposed by ansatz \cite{Acharya_S_1994} and now with several justifications \cite{Wu_1994, Kaniadakis_etal_1996, Zhou_2000, Acharya_S_2004}; see also \S\ref{Urn}.
\item {\it Cattani-Fernandes statistics}, derived by a combinatorics argument using quantum group theory \cite{Cattani_F_1982, Cattani_F_1983, Cattani_F_1984}.
\end{list}
Other intermediate statistics have also been proposed. Their main application has been to quantum particle systems, but curiously, only in the asymptotic limit $N \to \infty$.  Gentile statistics have also been applied to the analysis of socioeconomic and transport systems, again only in asymptotic form \citep[e.g.][]{Kapur_K_1992}.  

\begin{turnpage}
\begin{table*}[t]
\begin{ruledtabular}
\begin{tabular}{p{39pt} p{32pt} p{70pt} 
p{43pt} p{150pt} p{160pt} p{1pt}} 
Name & Scheme$^a$ & 
Statistical weight $\mathbb{W}$ &
Restriction & 
\raggedright{ Combinatorial entropy $H=N^{-1} \ln \mathbb{W}$ \eqref{eq:Boltz2} } & 
Inferred MaxProb distribution$^b$ 
\\ 
\hline \hline
Maxwell-Boltzmann &
D to D&
\raggedright {$\mathbb{W}_{MB}  = 
N!\prod\limits_{i = 1}^s {\dfrac{{g_i ^{n_i } }}{{n_i !}}}$} 
&
None&
\raggedright {$H_{MB}^{(N)}  = 
\sum\limits_{i = 1}^s \frac{1}{N} \Bigl\{ p_i \ln N! - \ln [(p_i N)!]  +  p_i N \ln g_i  \Bigr\} $}
&
\raggedright {$p_{MB,i}^\#  = \frac{1}{N} \Lambda ^{ - 1} \Bigl[ \frac{1}{N} \ln N!  + \ln g_i  - \lambda _0  - \sum\limits_{r=1}^R \lambda_r f_{ri}  \Bigr]$}
&\\
\cline{4-7} 
&
&
&
\raggedright {$N \to \infty$} &
$H_{MB}   =  - \sum\limits_{i = 1}^s {p_i \ln \dfrac{{p_i }}{{g_i }}} $
&
$p_{MB,i} ^* = g_i \exp \Bigl( - \lambda _0 ' - \sum\limits_{r=1}^R \lambda_r f_{ri} \Bigr) $
\\
\hline
Lynden-Bell &
\raggedright {D to D exclusive}&
\raggedright {$\mathbb{W}_{LB}  =  N! \prod\limits_{i = 1}^s {\dfrac{{g_i !}}{{n_i ! \, (g_i  - n_i )!}}}  
$} 
&
\raggedright{$n_i \in \{0,1\}$}&
\raggedright {$H_{LB}^{(N)}  = \sum\limits_{i = 1}^s \frac{1}{N} \Bigl\{ p_i \ln N! +  \ln \bigl[ (\alpha _i N)! \bigr] - \ln \bigl[ (p_i N)! \bigr] - \ln \bigl[ (\alpha _i N - p_i N)! \bigr] \Bigr\} $}
&
\raggedright {$p_{LB,i}^\#  = \frac{1}{N} \Lambda ^{ - 1} \Bigl[ \frac{1}{N} \ln N! + \Lambda (\alpha _i N - p_{LB,i} ^\#  N) - \lambda _0  - \sum\limits_{r=1}^R \lambda_r f_{ri}  \Bigr] $}
&\\
\cline{4-7} 
&
&
&
\raggedright {and $N \to \infty$} &
\raggedright {$H_{LB}  = \sum\limits_{i = 1}^s \Bigl\{  - (\alpha _i  - p_i )\ln (\alpha _i  - p_i ) + \alpha _i \ln \alpha _i  - p_i \ln p_i  \Bigr\} $} &
\raggedright {$p_{LB,i} ^* = \dfrac{\alpha _i }{\exp \Bigl( \lambda _0  + \sum\limits_{r=1}^R \lambda_r f_{ri} \Bigr)  + 1}$}
&\\
\hline
\hline
Bose-Einstein &
I to D&
\raggedright {$\mathbb{W}_{BE}  = \prod\limits_{i = 1}^s {\dfrac{{(g_i  + n_i  - 1)!}}{{(g_i  - 1)! \, n_i !}}}$}  &
None&
\raggedright {$H_{BE}^{(N)}  = \sum\limits_{i = 1}^s \frac{1}{N} \Bigl\{ \ln \bigl[ {(\alpha _i N + p_i N - 1)!} \bigr] - \ln \bigl[ {(\alpha _i N - 1)!} \bigr] - \ln \bigl[ {(p_i N)!} \bigr] \Bigr\}
 $}
&
\raggedright {$p_{BE,i}^\#  = \frac{1}{N} \Lambda ^{ - 1} \Bigl[ \Lambda (\alpha _i N + p_{BE,i}^\#  N - 1) - \lambda _0  - \sum\limits_{r=1}^R \lambda_r f_{ri}  \Bigr]
$}
&\\
\cline{4-7} 
&
&
&
\raggedright {$N \to \infty$} &
$H_{BE}  = \sum\limits_{i = 1}^s \Bigl \{ (\alpha _i  + p_i )\ln (\alpha _i  + p_i )  - \alpha _i \ln \alpha _i  - p_i \ln p_i  \Bigr\} $ 
&
$p_{BE,i} ^* = \dfrac{\alpha _i }{\exp \Bigl(\lambda _0  + \sum\limits_{r=1}^R \lambda_r f_{ri} \Bigr)  - 1}$
\\
\hline
Fermi-Dirac &
\raggedright {I to D exclusive}&
\raggedright {$\mathbb{W}_{FD}  = \prod\limits_{i = 1}^s {\dfrac{{g_i !}}{{n_i ! \, (g_i  - n_i )!}}}  
$} 
&
\raggedright{$n_i \in \{0,1\}$}&
\raggedright {$H_{FD}^{(N)}  = \sum\limits_{i = 1}^s \frac{1}{N} \Bigl\{ \ln \bigl[ (\alpha _i N)! \bigr] - \ln \bigl[ (p_i N)! \bigr] - \ln \bigl[ (\alpha _i N - p_i N)! \bigr] \Bigr\} $}
&
\raggedright {$p_{FD,i}^\#  =\frac{1}{N} \Lambda ^{ - 1} \Bigl[ \Lambda (\alpha _i N - p_{FD,i} ^\#  N) - \lambda _0  - \sum\limits_{r=1}^R \lambda_r f_{ri}  \Bigr] $}
&\\
\cline{4-7} 
&
&
&
\raggedright {and $N \to \infty$} &
\raggedright {$H_{FD}  = \sum\limits_{i = 1}^s \Bigl\{  - (\alpha _i  - p_i )\ln (\alpha _i  - p_i ) + \alpha _i \ln \alpha _i  - p_i \ln p_i  \Bigr\} $} 
&
\raggedright {$p_{FD,i} ^* = \dfrac{\alpha _i }{\exp \Bigl( \lambda _0  + \sum\limits_{r=1}^R \lambda_r f_{ri} \Bigr)  + 1}$}
&\\
\end{tabular}
\end{ruledtabular}
\caption{Statistical weights and inferred distributions for systems with distinguishable degenerate categories, with $\alpha_i=g_i/N$. 
$^a$Allocation of entities to categories, with D=distinguishable and I=indistinguishable (subcategories are of same type as categories). 
$^b$Obtained by maximising $H$ subject to constraints \eqref{eq:C0}-\eqref{eq:C1}.}
\label{tab:disting}
\end{table*}
\end{turnpage}

\section{\label{Indisting_Cat}Distinguishability of Categories}
By logical extension of \S\ref{Indisting_Ent}, we can also consider the allocation of (in)distinguishable entities to indistinguishable categories.  Despite the fact that indistinguishable categories are part of the ``folklore'' of combinatorics, and are included in published tables of the number of combinations or permutations of different allocation schemes (e.g.\ the ``twelve-fold way'') \cite{Riordan_1978, Fang_1985, Stanley_1986, Zwillinger_2003, Proctor_2006}, the entropy functions and most probable distributions of such systems have only recently been examined \cite{Niven_CTNext07}.  For convenience, we define:
\begin{list}{$\bullet$}{\topsep 2pt \itemsep 2pt \parsep 0pt \leftmargin 8pt \rightmargin 0pt \listparindent 0pt \itemindent 0pt}
\item {\it D:I statistics}, in which $N$ distinguishable entities are allocated to $s$ indistinguishable categories;
\item {\it I:I statistics}, in which $N$ indistinguishable entities are allocated to $s$ indistinguishable categories.
\end{list}
The D:I case has been examined for univariate, non-degenerate and equally degenerate categories \cite{Niven_CTNext07}, whilst the I:I case has not previously been examined. In the following, the non-degenerate forms of each statistic are discussed in detail, followed by their equally degenerate forms.

\subsection{\label{NonDegenI}Non-Degenerate D:I and I:I Statistics}
%
\begin{figure}[h]
\begin{center}
\setlength{\unitlength}{0.6pt}
  \begin{picture}(280,430)
   \put(25,215){\includegraphics[width=50mm]{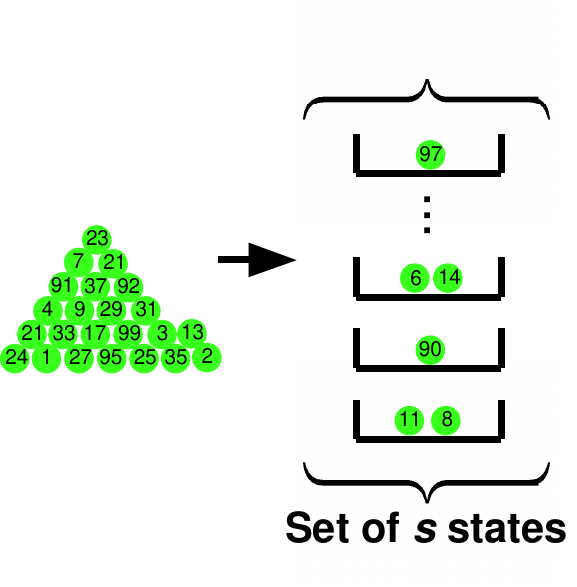} }
   \put(0,225){\small (a)}
   \put(25,-15){\includegraphics[width=50mm]{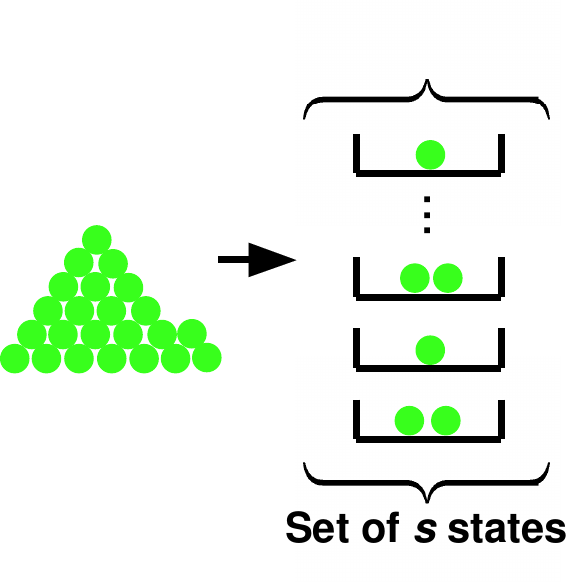} }
   \put(0,0){\small (b)}
  \end{picture}
\end{center}
\caption{Allocation schemes for non-degenerate indistinguishable categories: (a) D:I statistic and (b) I:I statistic.}
\label{fig:Indisting_Cat}
\end{figure}

Firstly examining the {\bf non-degenerate D:I statistic} illustrated in Figure \ref{fig:Indisting_Cat}(a), we denote the weight of each realization $\{n_i\}$ by:
\begin{equation} 
\mathbb{W}_{D:I} = \Ldbrace  \begin{matrix} N \\  {n_1, \dots, n_k, 0, \dots, 0}  \\  \end{matrix} \Rdbrace 
\label{eq:W_DtoI_def}
\end{equation}
where $k \le s$ is the number of filled categories $n_i >0$. By combinatorial enumeration of some simple examples, the following features emerge \cite{Niven_CTNext07}:
\begin{list}{$\bullet$}{\topsep 2pt \itemsep 2pt \parsep 0pt \leftmargin 8pt \rightmargin 0pt \listparindent 0pt \itemindent 0pt}
\item Unfilled categories do not affect the weight, i.e.\ \cite{Niven_CTNext07}:
\begin{equation} 
\Ldbrace  \begin{matrix} N \\  {n_1, \dots, n_k, 0, \dots, 0}  \\  \end{matrix} \Rdbrace  = \Ldbrace  \begin{matrix} N \\  {n_1, \dots, n_k}  \\  \end{matrix} \Rdbrace 
\label{eq:W_DtoI_zeros}
\end{equation}
\item Permutations of the occupancies are meaningless, e.g.\ $\{1,2,1\}$ and $\{1,1,2\}$ refer to the {\it same} realization \cite{Niven_CTNext07}. This is quite different to multinomial and quantum systems (\S\ref{Mult}-\ref{Indisting_Ent}), in which permuting the occupancies generates different realizations.
\end{list}
It can be shown that the weight is \cite{Niven_CTNext07}:
\begin{equation}
\begin{split}
\mathbb{W}_{D:I} 
&= \frac {N!} { \Bigl( \prod\limits_{i=1}^k  {n_i!} \Bigr) \Bigl( \prod\limits_{j=1}^N {r_j!} \Bigr) }
= \frac {N!} { \Bigl( \prod\limits_{i=1}^s  {n_i!} \Bigr) \Bigl( \prod\limits_{j=1}^N {r_j!} \Bigr) }
\end{split}
\label{eq:W_DtoI}
\end{equation}
where $r_j \ge 0$ is the {\it repetitivity}, or number of occurrences of integer $j$ in the realization $\{ n_i \}$ (without counting zeros), hence $\sum\nolimits_{j=1}^{N} r_j =k$.  Proof of \eqref{eq:W_DtoI} follows from the successive filling of cells \cite{Niven_CTNext07}.  The weight satisfies \cite{Abramowitz_S_1965, Char_2002, Niven_CTNext07}:
\begin{align}
\Bigl \{  \begin{matrix} N \\  k  \\  \end{matrix} \Bigr \} &= \sum\limits_{\begin{smallmatrix}  \text{all } \{n_i\} \\ \text{fixed } k \end{smallmatrix} }  \Ldbrace  \begin{matrix} N \\  {n_1, ..., n_k, 0, ..., 0}  \\  \end{matrix} \Rdbrace
\label{eq:thm1}
\\
\begin{split}
B(N,s) &= \sum\limits_{k=1}^s \Bigl \{  \begin{matrix} N \\  k  \\  \end{matrix} \Bigr \} = \sum\limits_{k=1}^s \hspace{4pt} \back \sum\limits_{\begin{smallmatrix}  \text{all } \{n_i\} \\ \text{fixed } k \end{smallmatrix} }  \! \Ldbrace  \begin{matrix} N \\  {n_1, ..., n_k, 0, ..., 0}  \\  \end{matrix} \Rdbrace
\end{split}
\label{eq:Bell}
\end{align}
where $\bigl \{  \begin{smallmatrix} N \\  k  \\  \end{smallmatrix} \bigr \}$ is a Stirling number of the second kind and $B(N,s)$ is an {incomplete Bell number}, equal to the total number of configurations \cite{Riordan_1978, Stanley_1986, Zwillinger_2003, Proctor_2006}. $B(N,s)$ reduces to the usual Bell number $B_N$ \cite{Zwillinger_2003} for $s=N$. 

Applying the combinatorial definition \eqref{eq:Boltz2} with $K=N^{-1}$ to \eqref{eq:W_DtoI} yields the non-asymptotic entropy \cite{Niven_CTNext07}:
\begin{align}
\begin{split}
H_{D:I}^{(N)} 
= \frac{1}{N} \sum\limits_{i=1}^s \Bigl( \frac{n_i}{N} \ln N! - \ln n_i! \Bigr) - \frac{1}{N} \sum\limits_{j=1}^{N} \ln r_j!
\end{split}
\label{eq:H_DtoI_exact}
\end{align}
where the $\ln N!$ term is brought inside the first sum using $\sum\nolimits_{i=1}^s n_i=N$. As evident, finding the asymptotic form or extremisation of \eqref{eq:H_DtoI_exact} requires careful handling of the $\{r_j\}$, and is therefore not as straightforward as in classical or quantum statistics. For $N \! \to \! \infty$ (hence $s \back \ll \back N$) and $n_i \! \to \! \infty, \forall i$, application of the Stirling approximation $\ln m! \approx m \ln m - m$ and the associated limits $r_{j \ne \infty} = 0$, $r_{\infty} = k$ gives, for $k \ll \infty$ \cite{Niven_CTNext07}:
\begin{equation}
\lim\limits_{\begin{smallmatrix} N \to \infty \\  n_i \to \infty, \forall i  \\  \end{smallmatrix}} 
H_{D:I}^{(N)} 
= - \sum\limits_{i = 1}^s {p_i \ln p_i }
\label{eq:H_DtoI_limit}
\end{equation}
$H_{D:I}^{(N)}$ thus converges to the Shannon entropy \eqref{eq:Shannon} under these conditions. Outside of these limits, e.g.\ for $s \gtrsim N$, \eqref{eq:H_DtoI_limit} does not apply, since it is critically dependent on $n_i \to \infty, \forall i$, not just on $N \to \infty$ \cite{Niven_CTNext07}. The D:I statistic thus differs substantially from the multinomial in its asymptotic properties.

We next examine the {\bf non-degenerate I:I statistic} (Figure \ref{fig:Indisting_Cat}(b)).  The weight can be denoted:
\begin{equation} 
\mathbb{W}_{I:I} = \Ldbrack  \begin{matrix} N \\  {n_1, \dots, n_k, 0, \dots, 0}  \\  \end{matrix} \Rdbrack 
\label{eq:W_ItoI_def}
\end{equation}

\begin{table*}[t]
\begin{tabular}{p{15pt} p{15pt} p{160pt} p{50pt} p{60pt} p{160pt} l}  
\hline
$N$	&$s$	&Non-degen. MB statistic only &&&Non-degen. MB and BE statistics \\
&&Actual MaxProb realization(s) $[n_i^\#]$	&$\mathbb{W}_{MB}$ (each)	&$\mathbb{P}_{MB}$ (each) &MeanProb realization $[\overline{n}_i]$ \\
\hline
1	&1	&[1]	&1	&1	&[1]\\
2	&2	&[1, 1]	&2	&0.5	&[1, 1]\\
3	&3	&[1, 1, 1]	&6	&0.222222	&[1, 1, 1]\\
4	&4	&[1, 1, 1, 1]	&24	&0.093750	&[1, 1, 1, 1]\\
5	&5	&[1, 1, 1, 1, 1]	&120	&0.038400	&[1, 1, 1, 1, 1]\\
10	&10	&[1, \dots, 1]	&3.63E+06	&3.629E-04	&[1, \dots, 1]\\
20	&20	&[1, \dots, 1]	&2.43E+18	&2.320E-08	&[1, \dots, 1]\\
30	&30	&[1, \dots, 1]	&2.65E+32	&1.288E-12	&[1, \dots, 1]\\
40	&40	&[1, \dots, 1]	&8.16E+47	&6.749E-17	&[1, \dots, 1]\\
50	&50	&[1, \dots, 1]	&3.04E+64	&3.424E-21	&[1, \dots, 1]\\
\hline
1	&3	&[1, 0, 0], [0, 1, 0], [0, 0, 1]	&1	&0.333333	&[1/3, 1/3, 1/3]\\
2	&3	&[1, 1, 0], [1, 0, 1], [0, 1, 1]	&2	&0.222222	&[2/3, 2/3, 2/3]\\
3	&3	&[1, 1, 1]	&6	&0.222222	&[1, 1, 1]\\
4	&3	&[1, 1, 2], [1, 2, 1], [2, 1, 1]	&12	&0.148148	&[4/3, 4/3, 4/3]\\
5	&3	&[1, 2, 2], [2, 1, 2], [2, 2, 1]	&30	&0.123457	&[5/3, 5/3, 5/3]\\
10	&3	&[3, 3, 4], [3, 4, 3], [4, 3, 3]	&4200	&0.071127	&[10/3, 10/3, 10/3]\\
20	&3	&[6, 7, 7], [7, 6, 7], [7, 7, 6]	&1.33E+08	&0.038151	&[20/3, 20/3, 20/3]\\
30	&3	&[10, 10, 10]	&5.55E+12	&0.026961	&[10, 10, 10]\\
40	&3	&[13, 13, 14], [13, 14, 13], [14, 13, 13]	&2.41E+17	&0.019853	&[40/3, 40/3, 40/3]\\
50	&3	&[16, 17, 17], [17, 16, 17], [17, 17, 16]	&1.15E+22	&0.016005	&[50/3, 50/3, 50/3]\\
\hline
\end{tabular}
\caption []{MaxProb and MeanProb realizations for non-degenerate MB and BE statistics,  subject to \eqref{eq:C0} (in part after \cite{Niven_CTNext07}).}
\label{tab:MB_BE_data}
\end{table*} 

\begin{table*}[t]
\begin{tabular}{p{15pt} p{15pt} p{130pt} p{47pt} p{48pt} p{200pt} l}  
\hline
$N$	&$s$	&Non-degenerate D:I statistic\\
&&\raggedright{Actual MaxProb realization(s) $\{n_i^\#\}$}	&$\mathbb{W}_{D:I}$ (each)	&$\mathbb{P}_{D:I}$ \text{  } (each) &MeanProb realization $\{\overline{n}_i\}$\\
\hline
1	&1	&\{1\}	&1	&1	&\{1\}\\
2	&2	&\{1, 1\}, \{2, 0\}	&1	&0.5	&\{1.5, 0.5\}\\
3	&3	&\{2, 1, 0\}	&3	&0.6	&\{2, 0.8, 0.2\}\\
4	&4	&\{2, 1, 1, 0\}	&6	&0.4	&\{2.333, 1.133, 0.467, 0.067\}\\
5	&5	&\{2, 2, 1, 0, 0\}	&15	&0.288462	&\{2.615, 1.462, 0.692, 0.212, 0.019\}\\
6	&6	&\{3, 2, 1, 0, 0, 0\}	&60	&0.295567	&\{2.842, 1.759, 0.916, 0.399, 0.079, 4.93E-03\}\\
7	&7	&\{3, 2, 1, 1, 0, 0, 0\}	&210	&0.239453	&\{3.058, 1.981, 1.166, 0.584, 0.185, 0.025, 1.14E-03\}\\
8	&8	&\{3, 2, 2, 1, 0, \dots, 0\}	&840	&0.202899	&\{3.245, 2.173, 1.417, 0.761, 0.325, 0.071, 7.00E-03, 2.42E-04\}\\
9	&9	&\{3, 2, 2, 1, 1, 0, \dots, 0\}	&3780	&0.178749	&\{3.419, 2.337, 1.643, 0.949, 0.477, 0.149, 0.024, 1.75E-03, 4.73E-05\}\\
10	&10	&\hspace{-3pt}$\begin{array}[t]{l} \{3, 2, 2, 1, 1, 1, 0, \dots, 0\}, \\ \{3, 2, 2, 2, 1, 0, \dots, 0\}, \\\{3, 3, 2, 1, 1, 0, \dots, 0\}, \\ \{4, 3, 2, 1, 0, \dots, 0, 0, 0\} \end{array}$	&12600	&0.108644	&\{3.576, 2.494, 1.827, 1.154, 0.629, 0.254, 0.058, 6.86E-03, 3.97E-04, 8.62E-06\}\\
20	&20	&\hspace{-3pt}$\begin{array}[t]{l} \{4, 3, 3, 2, 2, 2, 2, 1, 1, 0, \dots, 0\}, \\ \{4, 4, 3, 3, 2, 2, 1, 1, 0, \dots, 0\} \end{array}$	&1.83E+12	&0.035443	&\{4.677, 3.623, 2.999, 2.479, 2.046, 1.666, 1.169, 0.729, 0.395, 0.160, 0.046, 9.26E-03, \dots, 1.93E-14\}\\
30	&30	&\hspace{-3pt}$\begin{array}[t]{l} \{5, 4, 4, 3, 3, 3, 2, 2, 2, 1, 1, \\ 0, \dots, 0\} \end{array}$	&1.54E+22	&0.018214	&\{5.376, 4.330, 3.710, 3.244, 2.880, 2.495, 2.131, 1.858, 1.507, 1.078, 0.703, 0.406, 0.191, 0.069, 0.019, 4.17E-03, \dots, 5.15E-22, 1.18E-24\}\\
40	&40	&\hspace{-3pt}$\begin{array}[t]{l} \{5, 4, 4, 4, 3, 3, 3, 3, 2, 2, 2, 2, \\ 1, 1, 1, 0, \dots 0\} \end{array}$	&1.14E+33	&0.007265	&\{5.892, 4.848, 4.246, 3.797, 3.405, 3.093, 2.832, 2.508, 2.181, 1.946, 1.691, 1.333, 0.952, 0.627, 0.366, 0.180, 0.072, 0.023, 5.93E-03, \dots, 6.35E-36\}\\
50	&50	&\hspace{-3pt}$\begin{array}[t]{l} \{6, 5, 5, 4, 4, 4, 3, 3, 3, 3, 2, 2, \\ 2, 2, 1, 1, 0, \dots, 0\}	\end{array}$ &7.40E+44	&0.003986	&\{6.304, 5.262, 4.662, 4.225, 3.875, 3.542, 3.238, 3.016, 2.806, 2.516, 2.214, 1.996, 1.794, 1.505, 1.152, 0.815, 0.531, 0.307, 0.151, 0.062, 0.021, 6.00E-03, \dots, 5.38E-48\}\\
\hline
1	&3	&\{1, 0, 0\}	&1	&1	&\{1, 0, 0\}\\
2	&3	&\{1, 1, 0\}, \{2, 0, 0\}	&1	&0.5	&\{1.5, 0.5, 0\}\\
3	&3	&\{2, 1, 0\}	&3	&0.6	&\{2, 0.8, 0.2\}\\
4	&3	&\{2, 1, 1\}	&6	&0.428571	&\{2.429, 1.143, 0.429\}\\
5	&3	&\{2, 2, 1\}	&15	&0.365854	&\{2.805, 1.585, 0.610\}\\
6	&3	&\{3, 2, 1\}	&60	&0.491803	&\{3.246, 1.893, 0.861\}\\
7	&3	&\{3, 2, 2\}, \{4, 2, 1\}	&105	&0.287671	&\{3.682, 2.205, 1.112\}\\
8	&3	&\{3, 3, 2\}, \{4, 3, 1\}	&280	&0.255941	&\{4.077, 2.592, 1.331\}\\
9	&3	&\{4, 3, 2\}	&1260	&0.384029	&\{4.505, 2.903, 1.592\}\\
10	&3	&\{5, 3, 2\}	&2520	&0.256046	&\{4.927, 3.218, 1.855\}\\
20	&3	&\{8, 7, 5\}	&9.98E+07	&0.171680	&\{8.887, 6.582, 4.531\}\\
30	&3	&\{11, 10, 9\}	&5.05E+12	&0.147059	&\{12.717, 9.907, 7.376\}\\
40	&3	&\{15, 13, 12\}	&2.09E+17	&0.103236	&\{16.468, 13.236, 10.297\}\\
50	&3	&\{18, 17, 15\}	&1.02E+22	&0.085360	&\{20.162, 16.578, 13.261\}\\
\hline
\end{tabular}
\caption []{MaxProb and MeanProb realizations for non-degenerate D:I statistics, subject to \eqref{eq:C0} (in part after \cite{Niven_CTNext07}).}
\label{tab:DtoI_data}
\end{table*}

\begin{table*}[t]
\begin{tabular}{p{15pt} p{15pt} p{450pt} llllllll}  
\hline
$N$	&$s$	&Non-degenerate I:I statistic: MeanProb realization $\{\overline{n}_i\}$ \\
\hline
1	&1	&\{1\}\\
2	&2	&\{1.5, 0.5\}\\
3	&3	&\{2, 0.667, 0.333\}\\
4	&4	&\{2.4, 1, 0.4, 0.2\}\\
5	&5	&\{2.857, 1.143, 0.571, 0.286, 0.143\}\\
10	&10	&\{4.571, 2.262, 1.286, 0.786, 0.476, 0.286, 0.167, 0.095, 0.048, 0.024\}\\
20	&20	&\{7.384, 4.056, 2.603, 1.775, 1.239, 0.879, 0.625, 0.447, 0.316, 0.223, 0.155, 0.107, 0.072, 0.048, 0.030, 0.019, 0.011, 0.006, 3.19E-03, 1.59E-03\}\\
30	&30	&\{9.736, 5.628, 3.795, 2.714, 1.998, 1.496, 1.131, 0.860, 0.655, 0.499, 0.380, 0.288, 0.218, 0.164, 0.122, 0.091, 0.067, 0.049, \dots, 1.78E-04\}\\
40	&40	&\{11.826, 7.059, 4.903, 3.608, 2.735, 2.111, 1.647, 1.295, 1.023, 0.810, 0.642, 0.509, 0.403, 0.318, 0.251, 0.198, 0.155, 0.121, \dots, 2.68E-05\}\\
50	&50	&\{13.736, 8.390, 5.947, 4.462, 3.450, 2.717, 2.165, 1.739, 1.404, 1.138, 0.924, 0.752, 0.612, 0.498, 0.405, 0.329, 0.267, 0.216, 0.174, 0.141, 0.113, \dots, 4.90E-06\}\\
\hline
1	&3	&\{1, 0, 0\}\\
2	&3	&\{1.5, 0.5, 0\}\\
3	&3	&\{2, 0.667, 0.333\}\\
4	&3	&\{2.75, 1, 0.25\}\\
5	&3	&\{3.4, 1.2, 0.4\}\\
10	&3	&\{6.429, 2.643, 0.929\}\\
20	&3	&\{12.545, 5.409, 2.045\}\\
30	&3	&\{18.626, 8.187, 3.187\}\\
40	&3	&\{24.773, 10.955, 4.273\}\\
50	&3	&\{30.885, 13.731, 5.385\}\\
\hline
\end{tabular}
\caption []{MeanProb realizations for non-degenerate I:I statistics, subject to \eqref{eq:C0}.}
\label{tab:ItoI_data}
\end{table*} 


\begin{table*}[!]
\begin{tabular}{p{14pt} p{16pt} p{465pt} l}  
\hline
$N$	&$s$	&Non-degenerate I:I statistic: Total weighted occupancies $M_{I:I,i} = \sum\nolimits_{\nu} n_i^{(\nu)} \mathbb{W}_{I:I}^{(\nu)}$\\
\end{tabular}
\begin{tabular}{p{14pt} p{16pt} p{23pt} *{10}{p{22pt}} *{6}{p{18pt}}  *{4}{p{14pt}} }  
\hline
1	&1	&1\\ 
2	&2	&3, &1\\ 
3	&3	&6, &2, &1\\ 
4	&4	&12, &5, &2, &1\\ 
5	&5	&20, &8, &4, &2, &1\\ 
6	&6	&35, &16, &8, &4, &2, &1\\ 
7	&7	&54, &24, &13, &7, &4, &2, &1\\ 
8	&8	&86, &41, &22, &13, &7, &4, &2, &1\\ 
9	&9	&128, &61, &35, &20, &12, &7, &4, &2, &1\\ 
10	&10	&192, &95, &54, &33, &20, &12, &7, &4, &2, &1\\ 
11	&11	&275, &136, &80, &49, &31, &19, &12, &7, &4, &2, &1\\ 
12	&12	&399, &204, &121, &76, &48, &31, &19, &12, &7, &4, &2, &1\\ 
13	&13	&556, &284, &172, &109, &71, &46, &30, &19, &12, &7, &4, &2, &1\\ 
14	&14	&780, &407, &247, &160, &105, &70, &46, &30, &19, &12, &7, &4, &2, &1\\ 
15	&15	&1068, &560, &347, &225, &151, &101, &68, &45, &30, &19, &12, &7, &4, &2, &1\\ 
16	&16	&1463, &779, &484, &320, &215, &147, &100, &68, &45, &30, &19, &12, &7, &4, &2, &1\\ 
17	&17	&1965, &1050, &661, &439, &300, &206, &143, &98, &67, &45, &30, &19, &12, &7, &4, &2, &1\\ 
18	&18	&2644, &1432, &906, &608, &418, &292, &203, &142, &98, &67, &45, &30, &19, &12, &7, &4, &2, &1\\ 
19	&19	&3498, &1901, &1215, &820, &570, &400, &283, &199, &140, &97, &67, &45, &30, &19, &12, &7, &4, &2, &1\\ 
20	&20	&4630, &2543, &1632, &1113, &777, &551, &392, &280, &198, &140, &97, &67, &45, &30, &19, &12, &7, &4, &2, &1\\ 
\hline
\end{tabular}
\caption []{Total weighted occupancies for non-degenerate I:I statistics for $s=N$, subject only to \eqref{eq:C0}.}
\label{tab:ItoI_data2}
\end{table*} 

\begin{table*}[!]
\begin{tabular}{p{75pt} | p{205pt} | p{195pt} l}  
\hline
&Distinguishable Entities & Indistinguishable Entities\\
\hline
Distinguishable &{\bf Non-degenerate MB statistics} & {\bf Non-degenerate BE statistics} \\
Categories&MaxProb and MeanProb &MeanProb only; realizations equiprobable\\
&Highly symmetric &Highly symmetric\\
&\raggedright{Strongly asymptotic to uniform distribution}&\raggedright{Strongly asymptotic to uniform distribution}&\\
\hline
Indistinguishable & {\bf Non-degenerate D:I statistics} & {\bf Non-degenerate I:I statistics}\\
Categories&MaxProb and MeanProb &MeanProb only; realizations equiprobable\\
&Highly asymmetric &Highly asymmetric\\
&\raggedright{Slowly asymptotic to uniform distribution, $s \ll N$} & \raggedright{Non-asymptotic to uniform distribution, $s \ll N$}&\\
&Non-asymptotic for $s=N$ ? & Monotonic decreasing asymptote \eqref{eq:ItoI_asympt2} for $s=N$\\
\hline
\end{tabular}
\caption []{Properties of non-degenerate statistics, subject only to \eqref{eq:C0}.}
\label{tab:stats}
\end{table*} 

\noindent I:I statistics have many features in common with the D:I case, e.g.\ unfilled categories have no effect on the realization or weight, and permutations of the occupancies are meaningless.  However, by inspection it is readily seen that, in the non-degenerate case:
\begin{equation}
\begin{split}
\mathbb{W}_{I:I} =1
\end{split}
\label{eq:W_ItoI}
\end{equation}
In other words, each realization is equiprobable, rendering the MaxProb principle ineffective; non-degenerate BE and FD statistics also exhibit this property (Table \ref{tab:disting}). Such systems must be examined using the MeanProb measure \eqref{eq:MeanProb} (in effect, a weighted average MaxProb). For completeness, it can also be shown that:
\begin{align}
&\mathcal{P}_k(N) = \back \sum\limits_{\begin{smallmatrix}  \text{all } \{n_i\} \\ \text{fixed } k \end{smallmatrix} }  \Ldbrack  \begin{matrix} N \\  {n_1, ..., n_k, 0, ..., 0}  \\  \end{matrix} \Rdbrack 
= \back \sum\limits_{\begin{smallmatrix}  \text{all } \{n_i\} \\ \text{fixed } k \end{smallmatrix} } \back 1
\label{eq:partition}
\\
\begin{split}
&\mathcal{P}(N) = \sum\limits_{k=1}^s \mathcal{P}_k(N) = \sum\limits_{k=1}^s \hspace{4pt} \back \sum\limits_{\begin{smallmatrix}  \text{all } \{n_i\} \\ \text{fixed } k \end{smallmatrix} } \! \Ldbrack  \begin{matrix} N \\  {n_1, ..., n_k, 0, ..., 0}  \\  \end{matrix} \Rdbrack
\end{split}
\label{eq:cumpartition}
\end{align}
where $\mathcal{P}_k(N)$ is a partition number and $\mathcal{P}(N)$ a cumulative partition number \cite{Zwillinger_2003}; the latter gives the total number of configurations \cite{Riordan_1978, Stanley_1986, Zwillinger_2003, Proctor_2006}.

To consider some examples, the MaxProb (where possible) and MeanProb realizations of non-degenerate MB, BE, D:I and I:I systems subject only to the normalisation constraint \eqref{eq:C0}, calculated by enumeration of all configurations, are listed in Tables \ref{tab:MB_BE_data}-\ref{tab:ItoI_data} for various values of $s$ and $N$.  The MB and BE realizations are given as lists $[n_1, ..., n_s]$, whilst the D:I and I:I realizations are represented as ordered sets $\{n_1 \ge ... \ge n_s\}$ (the order is immaterial but convenient).   As evident:
\begin{list}{$\bullet$}{\topsep 2pt \itemsep 3pt \parsep 0pt \leftmargin 8pt \rightmargin 0pt \listparindent 0pt \itemindent 0pt}
\item The non-degenerate MB statistic (Table \ref{tab:MB_BE_data}) is highly symmetric, in that the entities try to spread as uniformly as possible over all available categories in both the MaxProb and MeanProb distributions. It is also strongly asymptotic, in that the MaxProb and MeanProb distributions converge rapidly to the uniform distribution, equivalent to the asymptotic distribution obtained by maximising the Shannon entropy \eqref{eq:Shannon}.
\item The non-degenerate BE statistic (Table \ref{tab:MB_BE_data}) is also highly symmetric and strongly asymptotic to a uniform distribution, as shown by its MeanProb distribution.
\item In contrast, the non-degenerate D:I statistic is highly asymmetric:\ its MaxProb distribution has a ``staircase'' appearance, in many cases cascading to a region of unoccupied cells, whilst the MeanProb distribution decreases monotonically but remains positive. For $s \back = \back N$, this statistic appears to be inherently non-asymptotic, with no obvious convergence of the MaxProb or MeanProb distributions to any function; they also differ significantly from each other.  For $s \back \ll \back N$ (illustrated by $s=3$), the MaxProb and MeanProb distributions converge slowly towards the uniform distribution, given by the Shannon asymptotic form \eqref{eq:H_DtoI_limit}.
\item The non-degenerate I:I statistic is also highly asymmetric,  even more so than the D:I case; its MeanProb distribution decreases monotonically but remains positive. It has no evident Shannon-like asymptotic convergence either for $s \back = \back N$ or $s \back \ll \back N$. However, for $s \back = \back N$ it does exhibit a curious asymptotic form, as revealed by the total weighted occupancies $M_{I:I,i} \back = \back \sum\nolimits_{\nu} n_{I:I,i}^{(\nu)} \mathbb{W}_{I:I}^{(\nu)}$ in Table \ref{tab:ItoI_data2}; these, divided by the total weights $\sum\nolimits_{\nu} \mathbb{W}_{I:I}^{(\nu)} \back = \back \mathcal{P}(N)$, give the MeanProb distribution \eqref{eq:MeanProb}.  As shown, $M_{I:I,i}$ converges as $N \to \infty$ to the sequence $1, 2, 4, 7, 12, 19, 30, 45, 67, 97, 139, ...$ rearranged in descending order; this is simply the sum (from zero) of partition numbers \cite{Riordan_1968, ATT_2008}. This leads to the following:

\vspace{3pt}
{\it Conjecture}: For $s=N$, the numerator of the MeanProb distribution of the non-degenerate I:I statistic satisfies:
\begin{equation}
\lim\limits_{N \to \infty} M_{I:I,i} = \lim\limits_{N \to \infty} \sum\limits_{\nu} n_i^{(\nu)} \mathbb{W}_{I:I}^{(\nu)} = \sum\limits_{\alpha=0}^{N-i} \mathcal{P}(\alpha)
\label{eq:ItoI_asympt1}
\end{equation}

{\it Corollary}: For $s=N$, the MeanProb distribution of the non-degenerate I:I statistic satisfies:
\begin{equation}
\lim\limits_{N \to \infty} \overline{n}_{I:I,i} = \frac {\sum\limits_{\alpha=0}^{N-i} \mathcal{P}(\alpha) } {\mathcal{P}(N)}
\label{eq:ItoI_asympt2}
\end{equation}

No attempt is made to prove these limits here. Convergence is quite rapid (valid at low $N$) towards the small end of the sequence ($i \to N$).
\end{list}
Non-degenerate D:I and I:I statistics therefore differ markedly from MB and BE statistics. Their properties are summarised in Table \ref{tab:stats}. For indistinguishable categories, it is seen that asymmetry is inherent, whilst for distinguishable categories, asymmetry can only arise from a non-uniform degeneracy and/or the imposition of moment constraints \eqref{eq:C1}.

\subsection{\label{DegenI}Equally Degenerate D:I and I:I Statistics}

Now consider {\bf equally degenerate D:I statistics}, in which each category $i$ contains $g$ equiprobable indistinguishable subcategories. The weight can be denoted \cite{Niven_CTNext07}:
\begin{equation}
\mathbb{W}_{D:I(g)} = \Ldbrace  \begin{matrix} N \\  {n_1, \dots, n_s}  \\  \end{matrix} \Rdbrace _{(g)} = \Ldbbrace  \begin{smallmatrix} & N &\\  n_{11}, &\dots, &n_{s1} \\ {\scriptstyle \vdots} & &\vdots  \\ n_{1g}, &\dots, &n_{sg}  \\  \end{smallmatrix} \Rdbbrace
\end{equation}
where $n_{im}$ is the occupancy of subcategory $m$ (hence $\sum\nolimits_{m=1}^g n_{im} = n_i$). Again $k \le s$ is the number of filled categories. The weight and entropy are obtained as \cite{Niven_CTNext07}:
\begin{align}
& \mathbb{W}_{D:I(g)} 
= \frac {N!} { \Bigl( \prod\limits_{i=1}^k  {n_i!} \Bigr) \Bigl( \prod\limits_{j=1}^N {r_j!} \Bigr) }  \prod\limits_{i=1}^k \hspace{4pt}  \sum\limits_{\gamma=1}^{\min(g,n_i)} \Bigl \{  \begin{matrix} n_i \\  \gamma  \\  \end{matrix} \Bigr \} 
\label{eq:W_DtoIg}
\\
\begin{split}
& H_{D:I(g)}^{(N)} 
= \frac{1}{N} \sum\limits_{i=1}^s \biggl( \frac{n_i}{N} \ln N!   - \ln n_i!  + \ln \sum\limits_{\gamma=1}^{\min(g,n_i)} \Bigl \{  \begin{matrix} n_i \\  \gamma  \\  \end{matrix} \Bigr \}  \biggr ) \\
& \qquad - \frac{1}{N} \sum\limits_{j=1}^{N} \ln r_j! 
\end{split}
\raisetag{22pt}
\label{eq:H_DtoIg_exact}
\end{align}
where $\gamma$ is an index of filled subcategories. Details of the derivation of \eqref{eq:W_DtoIg} are given in \cite{Niven_CTNext07}. For $N \to \infty$, $n_i \to \infty, \forall i$, $\lim\nolimits_{m \to \infty} \bigl \{  \begin{smallmatrix} m \\  a  \\  \end{smallmatrix} \bigr \} = a^m/a!$ \cite{Jordan_1947}, $r_{j \ne \infty} = 0$ and $r_{\infty} = k \ll \infty$, \eqref{eq:H_DtoIg_exact} converges to the MB-like entropy $H_{D:I(g)} =- \sum\nolimits_{i = 1}^s {p_i \ln {p_i}/{\gamma_i^\#} }$, where $\bigl \{  \begin{smallmatrix} n_i \\  \gamma_i^\#  \\  \end{smallmatrix} \bigr \}$ is the dominant term in the sum over $\gamma$.  Outside these limits, this asymptotic form is not obtained. 

For {\bf equally degenerate I:I statistics}, the weight can be denoted by:
\begin{equation}
\mathbb{W}_{I:I(g)} = \Ldbrack  \begin{matrix} N \\  {n_1, \dots, n_s}  \\  \end{matrix} \Rdbrack _{(g)} = \Ldbbrack  \begin{smallmatrix} & N &\\  n_{11}, &\dots, &n_{s1} \\ {\scriptstyle \vdots} & &\vdots  \\ n_{1g}, &\dots, &n_{sg}  \\  \end{smallmatrix} \Rdbbrack
\end{equation}
By enumeration of numerous examples, it can be established that the weight is given by:
\begin{align}
& \mathbb{W}_{I:I(g)} 
=  \prod\limits_{j =1}^{n_1} \; \al \Bigl( \sum\limits_{\gamma=1}^{\min(g,j)} \mathcal{P}_\gamma(j) \Bigr) ^{r_j}
\label{eq:W_ItoIg}
\end{align}
where $\al (a+b+...)^m$ is the {Wronski aleph function} \cite{Wronski_1811, Pragacz_2007}, a {combinatorial polynomial} or complete symmetric function \cite{Macdonald_1979, Lascoux_2003} given by a multinomial expansion with its coefficients omitted.  For example, consider:
\begin{align*}
(a+b)^2 &=a^2+2ab+b^2\\
(a+b)^3 &=a^3+3a^2b+3ab^2+b^3\\
(a+b)^m &=\sum\limits_{t=0}^m \bigl( \begin{smallmatrix} m \\  t \\  \end{smallmatrix} \bigr) a^t b^{m-t}
\end{align*}
where $\bigl( \begin{smallmatrix} m \\  t \\  \end{smallmatrix} \bigr)$ is the binomial coefficient. The Wronski forms are:
\begin{align*}
\al(a+b)^2 &=a^2+ab+b^2\\
\al(a+b)^3 &=a^3+a^2b+ab^2+b^3\\
\al(a+b)^m &=\sum\limits_{t=0}^m a^t b^{m-t}
\end{align*}
and in general:
\begin{align}
\al \biggl(\sum\limits_{\gamma=1}^{\Gamma} a_{\gamma} \biggr)^m= \sum\limits_{t_1, t_2, ..., t_{\Gamma}}  a_1^{t_1} a_2^{t_2} ... a_{\Gamma}^{t_{\Gamma}},
\label{eq:Wronski_aleph}
\end{align}
the sum taken over all permutations of $t_{\gamma} \ge 0$ which satisfy $\sum\nolimits_{\gamma=1}^{\Gamma} t_{\gamma}=m$. Proof of \eqref{eq:W_ItoIg} again proceeds from the successive filling of subcategories. An upper bound for the weight is given by the product, over all filled categories, of the number of subrealizations of $n_i$ entities in $\gamma$ subcategories; from \eqref{eq:partition}, the latter is given by:
\begin{alignat}{1}
\mathcal{P}_\gamma(n_i) = &\back \sum\limits_{\begin{smallmatrix}  \text{all } \{n_{im}\} \\ \text{fixed } \gamma \end{smallmatrix} }  \Ldbrack  \begin{matrix} n_i \\  {n_{i1}, ..., n_{i\gamma}, 0, ..., 0}  \\  \end{matrix} \Rdbrack 
= \back \sum\limits_{\begin{smallmatrix}  \text{all } \{n_{im}\} \\ \text{fixed } \gamma \end{smallmatrix} } \back 1 
\label{eq:subpartition}
\\
\text{whence:} \qquad &\mathbb{W}_{I:I(g)} 
\le  \prod\limits_{i =1}^{k} \Bigl( \sum\limits_{\gamma=1}^{\min(g,n_i)} \mathcal{P}_\gamma(n_i) \Bigr) 
\label{eq:W_ItoIg_upper}
\end{alignat}
The product of $\sum\nolimits_{\gamma=1}^{\min(g,n_i)} \mathcal{P}_\gamma(n_i)$ terms must then be modified to account for multiple occurrences of the same subrealization(s) in {\it different} categories, which are indistinguishable. This is achieved using the Wronski aleph instead of a polynomial product, whereupon \eqref{eq:W_ItoIg_upper} yields \eqref{eq:W_ItoIg} $\square$. 

The form of \eqref{eq:W_ItoIg}, based on the integers $j$ rather than occupancies $n_i$, is not very amenable for derivation of a combinatorial entropy function; further work is needed to determine if a more suitable form exists. In its absence, the MaxProb and MeanProb distributions can always be calculated using \eqref{eq:W_ItoIg} by enumeration of all realizations.

To this point, we have examined the effect of different features of ``ball-in-box'' allocation schemes (Figure \ref{fig:defs}), including system size (non-asymptotic effects), various types of degeneracy, (in)distinguishability of the balls or boxes and occupancy restrictions. Many more choices are possible, e.g.\ how the configurations should be amalgamated into realizations, other occupancy restrictions such as non-empty cells, ordered occupancies, mixtures of distinguishability types, etc \cite{Riordan_1978, Fang_1985, Stanley_1986, Zwillinger_2003, Proctor_2006}. Most of these options have not been examined from an entropic (inferential) perspective, and warrant further detailed investigation. 

\section{\label{Urn}Urn Models}
\begin{figure}[h]
\begin{center}
\setlength{\unitlength}{0.6pt}
  \begin{picture}(240,200)
   \put(0,-10){\includegraphics[width=45mm]{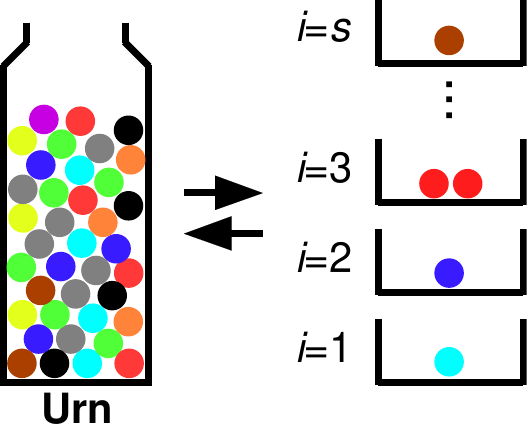} }
  \end{picture}
\end{center}
\caption{Urn model representation of a probabilistic system.}
\label{fig:urn}
\end{figure}

We now consider the use of {\it urn models} - related to but distinct from ``ball-in-box'' models - for the mathematical representation of probabilistic systems. Urn models have a long history, being employed by Jacob Bernoulli and Laplace \cite{Jaynes_2003}, and occupying the attention of many traditional statisticians during the 20th century \citep[e.g.][]{Feller_1957, Johnson_K_1977, Jensen_1985, Berg_1988}. A simple example is represented in Figure \ref{fig:urn}, in which balls are drawn from an urn containing a total of $M$ balls, made up of $m_i$ balls of the $i$th colour, for $i=1,...,s$.  A ball is drawn in accordance with some rule, recorded, and then returned to the urn and/or the urn modified in some manner; the sampling is repeated until a sample of $N$ balls, consisting of $n_i$ of each colour, is obtained \citep[c.f.][]{Jensen_1985, Berg_1988}. The urn model is used to generate the probability distribution $\mathbb{P}$ of the sampling scheme and, usually, its asymptotic behaviour ($M \to \infty$ and/or $N \to \infty$) is examined. Many extraordinarily complicated urn models have been devised, involving the conditional drawing and/or replacement of ball(s) from a single or multiple urns \cite{Feller_1957, Johnson_K_1977}.

Although a very old device, the new perspective here is that urn models generate a governing probability $\mathbb{P}$ which can be converted, by Boltzmann's principle \eqref{eq:Boltz3}, to a cross-entropy function \eqref{eq:Boltz3}.  One can then apply the tools of probabilistic inference, such as the MaxProb and MeanProb principles defined in \S\ref{Defs}, to infer the properties of the system. Surprisingly few physicists, mathematicians or information theorists have exploited this technique, despite Boltzmann's principle being over 130 years old \cite{Boltzmann_1877}.  Although it does simplify the calculations, it is not necessary that the system be asymptotic; furthermore, by the use of modern-day optimisation and numerical methods, many types of systems can be examined, such as those in which $\mathbb{P}$ is not in closed form.  Many quite complicated probabilistic systems involving conditional probabilities - e.g.\ Markovian or non-Markovian chains, random walks, networks, transport systems and games - can therefore be analysed in this manner.

It is known that MB, BE and FD statistics (\S\ref{Indisting_Ent}) can be constructed by simple urn models, respectively involving sampling with replacement, double replacement or without replacement, in the asympotic limits $M \! \to \! \infty, N \! \to \! \infty$ and $N/M \! \to \! \beta$ \cite{Jensen_1985, Berg_1988}.  Two recent studies \cite{Grendar_N_2007, Niven_G_MBBEFD_08} have extended these scenarios using the P\'olya urn model, in which the ball is returned after each draw and $c$ balls of the same colour are also added \cite{Eggenberger_P_1923, Polya_1931, Johnson_K_1997}:
\begin{equation}
\mathbb{P}_{Polya} = \frac{N!}{\prod\limits_{i=1}^s n_i!} 
\prod\limits_{i=1}^s \frac{m_i (m_i+c) \dots (m_i+(n_i-1)c)} {M(M+c) \dots (M+(N-1)c)},
\label{eq_Polya_distrib}
\end{equation}  
This is a closed-form example of ``neither independent nor identically distributed'' sampling, since the probability of drawing a ball of colour $i$ changes (conditionally) during sampling.  Eqs.\ \eqref{eq:Boltz3} and \eqref{eq_Polya_distrib} were then used to derive the P\'olya cross-entropy function. This includes MB, BE and FD statistics as special cases, and in general gives rise to the Acharya-Swamy intermediate statistic \cite{Niven_G_MBBEFD_08}. It is also shown that extremisation of the Kullback-Leibler function \eqref{eq:KL}, in a P\'olya system, infers a distribution which asymptotically vanishes and is therefore unrepresentative of the system \cite{Grendar_N_2007}. 

\section{\label{Graphs}Graphical Systems}
Finally, we consider systems which can be represented in graphical form. Graph theory is one of the mainstays of modern-day combinatorics, and there are few probabilistic systems which cannot be represented in this manner.  As well as graphs (formally defined below), a wide range of specialist concepts are available, including trees, networks, posets, cycles, chains, lattices and necklaces \citep[e.g.][]{Riordan_1978, Stanley_1986, Zwillinger_2003}.  As with urn models, the insight here is the ability to infer the ``typical'' properties of the system, for which the MaxProb and (possibly) the MeanProb principles are eminently suited. These may involve the derivation of an entropy or cross-entropy function, for extremisation subject to the constraints on the system. Curiously, however, few combinatorial or graph-theoretical studies invoke an entropy concept or seek the most probable realization of the system; most published studies which consider the graph entropy (defined below) stem from information theory \citep[e.g.][]{Korner_1973, Korner_L_1973, Korner_O_1998, Korner_S_2000, Simonyi_2000} .

We first define several terms \cite{Korner_1973, Korner_L_1973, Korner_O_1998, Korner_S_2000, Simonyi_2000, Zwillinger_2003, Abbass_2008}:
\begin{list}{$\bullet$}{\topsep 2pt \itemsep 2pt \parsep 0pt \leftmargin 8pt \rightmargin 0pt \listparindent 0pt \itemindent 0pt}
\item The {\it non-Cartesian product} of two sets $A$ and $B$ is given by $A \acute{\times} B=\{\{a,b\}| a \in A, b \in B\}$, i.e.\ the set of unordered pairs $\{a,b\}$ taken without repetition. 
\item An {\it undirected graph} is the ordered triple $G=(V,E,\psi)$, consisting of a non-empty finite set of vertices $V=\{v_i\}$, a finite set of edges $E=\{e_j\}$ with $E \cap V = \varnothing$ and a function $\psi: E \to V \acute{\times} V$. In other words, $\psi$ maps edges $e_i$ to a pair of vertices $\{v_{j},v_{k}\}$, without regard to order.
\item A {\it simple graph} is an undirected graph without single-node loops $e_i \to \{v_{j},v_{j}\}$ or multiple edges $e_i=e_t$.
\item A {\it complete graph} is a simple graph in which $\psi$ is surjective, i.e.\ all pairs of vertices have an edge. 
\item Two {\it complementary} graphs $G$ and $\overline{G}$ have the same vertex set $V$ and disjoint edge sets $E$ and $\overline{E}$, such that $\psi: E \cup \overline{E} \to V$ gives a complete graph. 
\item A {\it colouring} or {\it proper colouring} of a graph $G$ is a partition of the vertex set $V$ into edge-independent disjoint sets ({\it colour classes}), 
such that every edge joins vertices in two different colour classes.
\item The {\it chromatic number} $\chi(G)$ of a graph $G$ is the smallest number of classes in any colouring of $G$.
\item The {vertex packing polytope} $VP(G)$ of a graph $G$ is the convex hull of the characteristic vectors of stable sets of $G$ \cite{Simonyi_2000}. 
\end{list}
It is also possible to consider {\it directed graphs} or {\it digraphs}, in which each edge has a direction \cite{Zwillinger_2003, Abbass_2008}; these are not examined further here. 

\begin{figure}[t]
\begin{center}
\setlength{\unitlength}{0.6pt}
  \begin{picture}(120,150)
   \put(0,-25){\includegraphics[width=30mm]{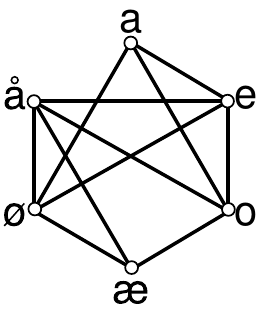} }
  \end{picture}
\end{center}
\caption{Graph of several Danish letters.}
\label{fig:Danish}
\end{figure}

The {\it graph entropy} concept follows from consideration of communications signals of length $N$, consisting of letters $v_j \in V$ from an alphabet $V$, represented as vertices of a graph. If the letters are considered distinguishable, they are made adjacent (joined by an edge). As an example, consider the Danish vowels in Figure \ref{fig:Danish}, which if scanned by English-language optical character recognition software, may exhibit the distinguishability relations shown. Its chromatic number $\chi=3$. The graph entropy of a simple graph $G$ on the vertex set $V=\{v_1,...,v_s\}$, with corresponding probability distribution $P=\{p_1,...,p_s\}$, is then defined as \cite{Korner_1973, Korner_L_1973, Korner_O_1998}
\begin{equation}
H(G,P) = \limsup\limits_{N \to \infty} \frac{1}{N} \log_2 \bigl(\chi(G_P^N)+1\bigr)
\label{eq:graph_entropy1}
\end{equation}
where $G_P^N$ implies a graph with distribution $P$ and signal length $N$.  A very different but more tractable definition, demonstrated to be equivalent, is \cite{Korner_S_2000, Simonyi_2000}:
\begin{equation}
H(G,P) = \min\limits_{a \in VP(G), a>0} \sum\limits_{i=1}^s p_i \log_2 \frac{1}{a_i}
\label{eq:graph_entropy2}
\end{equation}
A third definition of $H(G,P)$ is based on the mutual information \cite{Korner_O_1998, Simonyi_2000}. From a combinatorial perspective, the graph entropy enables the handling of categories with ``heterogeneous'' distinguishability, a superset of the D:I statistic analysed herein (\S\ref{Indisting_Cat}).  It also exhibits several interesting properties; e.g.\ the entropies of two complementary graphs are additive and equal to that of the complete graph \cite{Korner_S_2000, Simonyi_2000, Ramer_G_2006}, in some sense analogous to the additive nature of the thermodynamic entropy. The graph entropy is, however, exclusively asymptotic ($N \to \infty$). 

Substantially more research is required on the compatibility of the definition of graph entropy \eqref{eq:graph_entropy1}-\eqref{eq:graph_entropy2} with Boltzmann's principle, and on the application of probabilistic inference (e.g.\ the MaxProb principle) to systems represented in graphical form.

\section{\label{Concl}Conclusions}
This study examines {\it probabilistic systems} defined by $\Upsilon(U,C,\Psi)$, in which entities $u_m \in U$ are mapped to categories $c_{\bi} \in C$ by a probabilistic random variable $\Psi$; the resulting distinguishable {\it configurations} $\{U \to C\}$ are then grouped into {\it realizations} in accordance with some aggregation rule.  The {combinatorial} or {probabilistic} definitions of entropy $H$ and cross-entropy $D$, proportional respectively to the logarithm of the weight or probability of a specified realization \eqref{eq:Boltz2}-\eqref{eq:Boltz3} (``Boltzmann's principle''), are then considered.  These are defined so that extremisation of $H$ or $D$, subject to any constraints, always selects the ``most probable'' (MaxProb) realization(s) of the system \eqref{eq:MaxProb}.  Another useful measure of central tendency of a system is its mean-weighted (MeanProb) realization, the average of all realizations weighted by their weight or probability \cite{Grendar_G_2001}.  For multinomial systems, the combinatorial definitions \eqref{eq:Boltz2}-\eqref{eq:Boltz3} converge to the Shannon entropy or Kullback-Liebler cross-entropy in the asymptotic limit $N \to \infty$. However, as is made clear in this study, many systems may not be multinomial and/or may not have an asymptotic limit. Such systems cannot meaningfully be analysed with $D_{KL}$ or $H_{Sh}$, but can be analysed directly by MaxProb and/or MeanProb. This is illustrated by several examples, including (a) non-asymptotic systems; (b) systems with indistinguishable entities (quantum statistics); (c) systems with indistinguishable categories; (d) systems represented by urn models, such as ``neither independent nor identically distributed'' (ninid) sampling; and (e) systems representable in graphical form, such as decision trees and networks. Particular attention is devoted to (c), especially to analysis of the I:I statistic, including (i) identification of an asymptotic form of its non-degenerate MeanProb realization, and (ii) derivation of its non-degenerate statistical weight, in terms of partition numbers, coding parameters and the Wronski aleph function. The potential for significant new research, especially in (d) and (e), is also highlighted.

It is shown that the Boltzmann principle \eqref{eq:Boltz2}-\eqref{eq:Boltz3} leads to many different entropy or cross-entropy measures for different combinatorial systems, united by a common (MaxProb) principle \eqref{eq:MaxProb} founded in probability theory.  In contrast, the Shannon and Kullback-Leibler functions of information theory - which are often claimed to be universal measures of uncertainty applicable to all probabilistic systems \cite{Jaynes_1957, Jaynes_1963, Kapur_K_1992, Jaynes_2003} - do not have such a universal foundation. Indeed, in many systems, the distribution inferred by the Shannon or Kullback-Leibler functions can be shown to be unrepresentative of the system \cite{
Grendar_N_2007, Niven_2005, Niven_2006, Niven_CTNext07, Niven_non-asymp08, Niven_G_MBBEFD_08, Bose_1924, Einstein_1924, Einstein_1925, Fermi_1926, Dirac_1926, Brillouin_1927, Brillouin_1930, Tolman_1938, Davidson_1962}.  The combinatorial definition of entropy (Boltzmann's principle) is therefore of fundamentally greater importance, for the purpose of inferring the properties of a probabilistic system, than the definitions adopted in information theory.

\begin{acknowledgments}
The author thanks the organisers of the NEXT-$\Sigma\Phi$ 2008 conference in Kolymbari, Crete, Greece, at which this work was presented; the European Commission for financial support by a Marie Curie Incoming International Fellowship; The University of New South Wales for travel support; the participants of the Facets of Entropy workshop, Copenhagen, October 2007, expecially Marian Grendar, Bjarne Andresen, Ali Ghaderi, Flemming Tops\o e and Sergio Verdu, for interesting discussions; and Marian Grendar, Arthur Ramer and Hussein Abbass for guidance on graph theory.
\end{acknowledgments}


\bibliographystyle{aipproc}   

\end{document}